\documentclass[]{interact}

\usepackage{epstopdf}
\usepackage[caption=false]{subfig}

\usepackage{amsmath,amssymb,amsfonts}
\usepackage{algorithmic}
\usepackage{algorithm}
\usepackage{graphicx}
\usepackage{textcomp}
\usepackage{listings,jvlisting}

\usepackage[numbers,sort&compress]{natbib}
\bibpunct[, ]{[}{]}{,}{n}{,}{,}
\renewcommand\bibfont{\fontsize{10}{12}\selectfont}
\makeatletter
\def\NAT@def@citea{\def\@citea{\NAT@separator}}
\makeatother

\theoremstyle{plain}

\theoremstyle{definition}

\theoremstyle{remark}

\begin{document}


\title{Choice of Parallelism: Multi-GPU Driven Pipeline for Huge Academic Backbone Network}

\author{
\name{Ruo Ando\textsuperscript{a}\thanks{CONTACT Ruo Ando Email: ruo@nii.ac.jp}, Youki Kadobayashi \textsuperscript{b} and Hiroki Takakura\textsuperscript{a}}
\affil{\textsuperscript{a}National Institute of Informatics, 2-1-2 Hitotsubashi, Chiyoda-ku, Tokyo 101-8430 Japan \textsuperscript{b}Nara Institute of Science and Technology, Graduate School of Science and Technology, 8916-5 Takayama, Ikoma, Nara 630-0192, Japan}
}

\maketitle

\begin{abstract}
Science Information Network (SINET) is a Japanese academic backbone network for more than 800 research institutions and universities. 
In this paper, we present a multi-GPU-driven pipeline for handling huge session data of SINET.
Our pipeline consists of ELK stack, multi-GPU server, and Splunk. 
A multi-GPU server is responsible for two procedures: discrimination and histogramming. 
Discrimination is dividing session data into ingoing/outgoing with with subnet mask calculation and network address matching. 
Histogramming is grouping ingoing/outgoing session data into bins with map-reduce. 
In our architecture, we use GPU for the acceleration of ingress/egress discrimination of session data. 
Also, we use a tiling design pattern for building a two-stage map-reduce of CPU and GPU. 
Our multi-GPU-driven pipeline has succeeded in processing huge workloads of about 1.2 to 1.6 billion session streams (500GB-650GB) within 24 hours.
\end{abstract}



\maketitle

\section{Introduction}
From the dawn of the Internet era, monitoring is a fundamental part of network management. As recent years witnessed tremendous progress on high-speed networks, monitoring has become more and more complex. There have been many research efforts on managing and debugging large-scale networks. However, large-scale managing networks face more and more new challenges, which is also the case of academic backbone networks.

The Science Information Network (SINET) is a Japanese academic backbone network for more than 800 universities and research institutions. It connects many research facilities in seismology, space science, high-energy physics, nuclear fusion, computing science, and so on. It is now being used by over 2 million users and supports international research collaboration through international lines. In March 2019, the National Institute of Informatics (NII) built the world's first round-the-globe ultra-high-speed 100 Gbps academic communications network.

Since 2016, NII has been running a service of NII-SOCS (NII Security Operation Collaboration Services). Our team of NII-SOCS has deployed a security monitoring system consists of PA-7080, Elasticsearch, Splunk, and NVidia Multi-GPU server. We introduce our system and some operational experience of handling massive session data ranging from 400,000,000 to 800,000,000 per day in this talk. During the four years of 2016-2019, We have faced many challenges regarding several hosts, protocol proliferation, probe placement technologies, and security incident response.

Our system's main feature is multi-GPU accelerated time-series monitoring of SINET's huge session data in ELK Stack. The important insight we have obtained is that GPU's massively parallel processing power enables us to overcome conventional bottlenecks of multi-core CPU and horizontally scalable systems such as Hadoop, Spark, and so on. Three key technologies of our system are as follows:

\begin{itemize}


\item We propose a new application of direct NVMe access driven by sliced scroll for huge data pagination of Elasticsearch. In our application, each task assigned with a unique scroll ID directly accesses NVMe SSD with LBA calculation corresponding to the size of slice and JSON object. Besides, direct NVMe access is driven by a sliced scroll with the 1-1 correspondence between slice and task. By doing this, our application can eliminate READ latency tremendously while evading the implementation and deployment cost. In an experiment, we have adopted the SPDK perf tool and measured READ/WRITE latency in huge data pagination of Elasticsearch. Specifically, READ latency reduction leads to a significant performance improvement in huge JSON object parsing. It turned out that our SPDK based parser application can speed up the processing time compared with one of native Linux threads by ranging from x12.47 to x26.21 with the drastic reduction of reading latency.

\item Parallel CUDA Thrust API invocation on multi GPUs. We leverage CUDA Thrust API for rapid time-series generation from histogramming huge session data in Elasticsearch. CUDA Thrust is a high-performance API for histogramming session data with automatic parameters such as block and grid size. In our experience, Thrust usually outperforms the naive GPU kernel implementation without tidy parameter selection. Besides, Thrust API can run in parallel on a separate GPU. Approximately, we can enjoy the massive parallelism of  N (Pthreads) * M (CUDA threads) multithreading with the combination of traditional raw thread and GPU threads.

\item Highly concurrent container and raw threads. Grouping session data generate time-series data in our system into bins of milliseconds. Approximately, the granularity of histogramming is around 86,000,000 (60 \* 60 \* 24 \* 1000 = 86400000). For the rapid insertion of the fine-grained histogram data, we use a highly concurrent container, a concurrent vector of Intel Threading Building Block (TBB). We give up using a concurrent hash map, which suffers serious slowdown by lock contention. Instead, We adopt the concurrent vector. Key-value can be represented by using the defining namespace such as X1<timestamp>, X1<count> and X2<timestamp> and X2<bytes>. Also, hashmap Y1<timestamp, counts> and Y2<timestamp, bytes> are adopted.

\end{itemize}

To address the need for faster processing at scale, a multi-core CPU contains as many as 32-48 cores. But even the use of multi-core CPUs deployed in large clusters of servers can make sophisticated analytical applications unaffordable for all but a handful of organizations. An alternative and cost-effective way to address the compute performance bottleneck today is the general-purpose graphics processing unit (GPU). GPUs are capable of processing data up to 100 times faster than configurations containing CPUs alone. The reason for such a dramatic improvement in their massively parallel processing capabilities, with some GPUs containing nearly 6,000 cores upwards of 200 times more than the 32 to 48 cores found in today's most powerful CPUs.  
We leverage multi-GPU efficiency in terms of massive threads for speeding up the pipeline processing huge session data.

\section{Choice of PARALLELISM}

For handling huge session data on our pipeline, we adopt different kinds of parallelism. 
Our key finding is that choosing the appropriate parallelism level on each phase of the pipeline is important.
Figure 1 depicts our choice of parallelism with the constraints shown in the left part.
We have adopted four kinds of parallelism: machine level, process level, thread-level, and vector level.
In this section, we introduce Thrust Template Library (vector level),  Intel TBB (thread level), and POSIX Pthreads (thread level).


\begin{figure}[ht]
\centering
\includegraphics[scale=0.5]{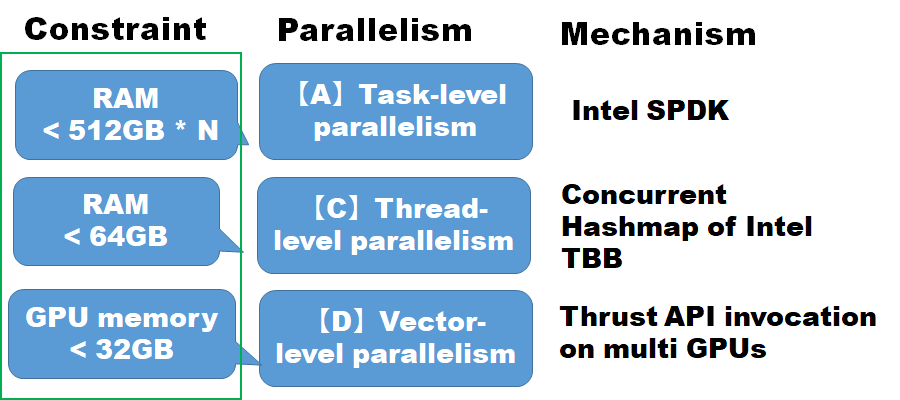}
\caption{Choice of parallelism for different constraints with different mechanisms.}
\end{figure}

\subsection{Vector level parallelism - CUDA THRUST}

Thrust is a C++ template library, both implementing and facilitating the implementation of parallel algorithms. Thrust's syntax resembles the Standard Template Library (STL), making it easy for seasoned C++ programmers to transition to parallel programming without going through the process of mastering complex tools like CUDA. Data in Thrust are stored in two types of vectors, which are functionally equivalent to the STL vector template class:

\begin{enumerate}
\item host\_vector. For data residing in host memory.
\item device\_vector. For data residing in device memory.
\end{enumerate}

The novelty about using these vector types is that data transfer between the host and the device or vice versa is implemented by overloading the assignment operator. Thrust provides efficient implementations for many important algorithms which can be used as building blocks. To problem solutions. These including sorting, scanning, subset selection, and reduction implementations. Not only does Thrust boost programmer productivity and program readability and maintenance, but it can also boost performance because it can adjust the execution configuration to the available GPU capabilities and resources. These algorithms fall into five categories.

\begin{enumerate}
\item Transformations.
\item Sorting and searching
\item Reductions
\item Scans/prefix-sums
\item Data management/manipulation
\end{enumerate}

Transformations operate an input sequence by applying a supplied operation on each element. Contrary to reduction,
the produced output is equal to size (in terms of granularity) to the input.

\subsection{Thread level parallelism - INTEL TBB}
Intel Threading Building Blocks (TBB) offers a rich and complete approach to expressing parallelism in a p++ program. It is a library that helps you leverage multi-core processor performance without having to be a threading expert. Threading Building Blocks is not just a threads-replacement library; it represents a higher-level, task-based parallelism that abstracts platform details and threading mechanisms for performance and scalability. Highly concurrent containers are very important because standard Template Library (STL) containers generally are not concurrency-friendly, and attempts to modify them concurrently can easily corrupt the containers. As a result, it is standard practice to wrap a lock (mutex) around STL containers to make them safe for concurrent access by letting only one thread operate on the container at a time. But that approach eliminates concurrency and thus is not conducive to multi-core parallelism. Intel Threading Building blocks provide highly concurrent containers that permit multiple threads to invoke a method simultaneously on the same container. At this time, a concurrent queue, vector, and hash map are provided.

\begin{enumerate}
\item concurrrent queue
\item vector
\item hashmap
\end{enumerate}

All of these highly concurrent containers can be used with this library, OpenMP, or raw threads.

\section{Coordinated Batch Processing}

To handle huge workloads of session data, we apply coordinated batch processing. Coordinated batch processing is essential as our workloads increase.
Coordinated batch processing is vital to pull multiple outputs back together in order to generate some sort of aggregated output.

The most canonical example of this aggregation is the reduced part of the map-reduce pattern.
It is easy to see that the map step is an example of sharding a work queue. The reduce step is an example of coordinated processing that eventually reduces a large number of outputs down to single aggregate response. However, there are many different aggregate patterns for batch processing, 
and this chapter discusses a number of them in addition to real-world applications.

Figure 2 depicts our coordinated batch processing. Before proceeding to the map-reduce phase, each workload (chunk) of session data 
is discriminated and marked as ingoing/outgoing.


\begin{figure}[ht]
\centering
\includegraphics[scale=0.6]{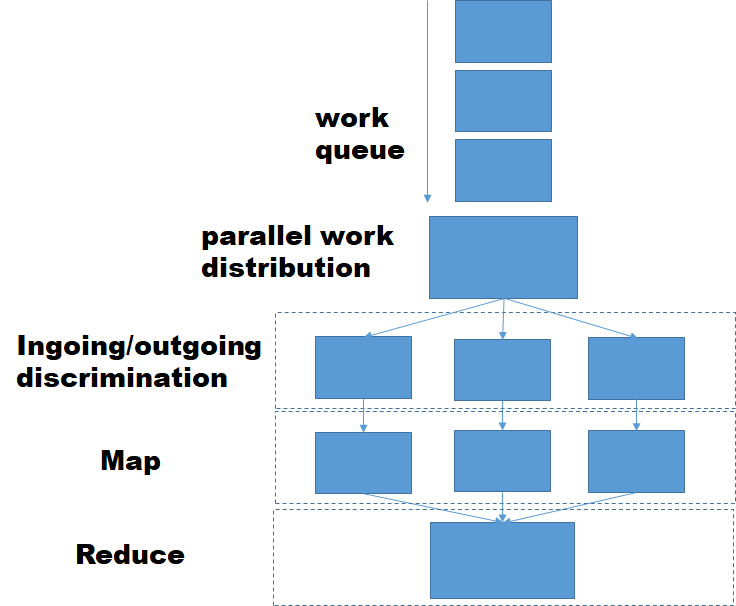}
\caption{Coordinated batch processing. Before the Map stage, each chunk in the work queue is discriminated against as ingoing/outgoing.}
\end{figure}

\subsection{Work Queue}
We adopt a work queue, which is the basic form of our pipeline. In Figure 3, work queue management runs in batch processing. 
In the view of concurrency, coordinated batch processing is a typical example because it can frequently happen regardless of The right side of Figure 3 shows discriminator threads by which each piece of work is independent of each other and can be processed without interactions. The main purpose of the work system's design is to ensure that each chunk of work is processed within a certain amount of time.

\subsection{Threads for traffic direction discrimination with CIDR MATCH}
In general, traffic direction discrimination is essential to preprocess. 
Before the stage of Map Reduce, session data should be discriminated against to outgoing or ingoing. 
Algorithm 1 depicts steps of direct discrimination. This algorithm is implemented in discriminator threads discussed in 
the previous section. 
The final goal of this algorithm is to mark ingoing/outgoing to each chunk of session data.
At line 6 and 7, we obtain a bit sequence of source I.P. address session data (S.B.) and SINET address range (C.B.).
Consequently, if S.D. is equal to C.B., the direction of the chunk is outgoing. 
Instead of matching S.D. to C.B., we use minus at line 8.
Finally, if the result equals to 0, we mark the chunk $outgoing$.
If the result equals 1, we mark the chunk $ingoing$.


\begin{figure}[ht]
\centering
\includegraphics[scale=0.5]{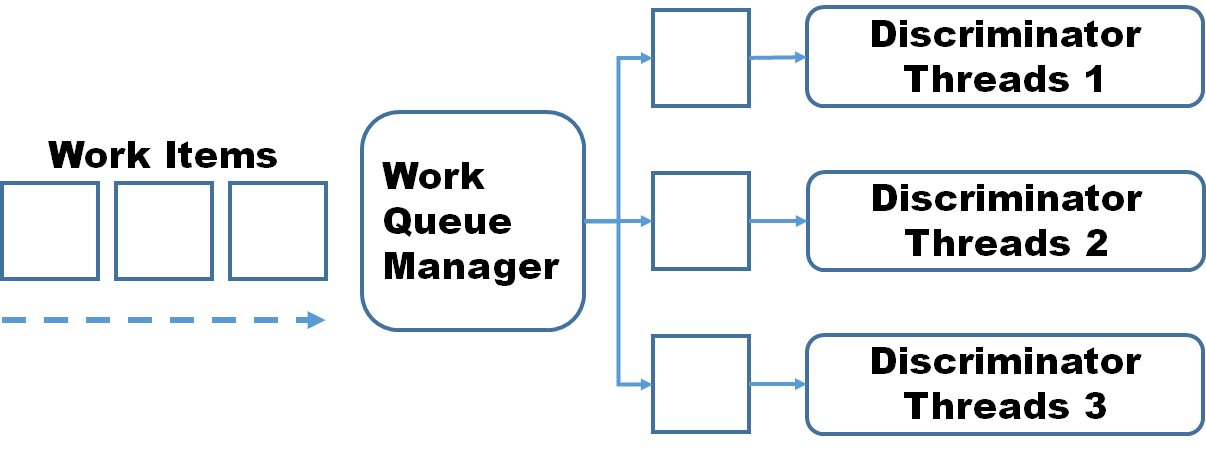}
\caption{The main purpose of the work system's design is to ensure that each chunk of work is processed within a certain amount of time.}
\end{figure}

\begin{algorithm}
  \caption{session ingoing/outgoing discrimination}
  \label{alg1}
  \begin{algorithmic}[1]
    \renewcommand{\algorithmicrequire}{\textbf{Input: hunks of session data}}
    \renewcommand{\algorithmicensure}{\textbf{Output:}}
    \STATE INPUT: chuncks of session data / CIDR list
    \STATE OUTPUT: chuncks of ingoing session data 
    \WHILE{chunk in workqueue is not EMPTY}
    \STATE CIDR notation = Y.Y.Y.Y\slash Z 
    \STATE source IP notation = X.X.X.X\slash Z   
    \STATE SB  = bitmask (X.X.X.X, Z)
    \STATE CB = bitmask(Y.Y.Y.Y, Z)
    \STATE result = SB - CB
    \IF{$ result == 0 $}
    \STATE mark(chunk, outgoing) 
    \ELSE
    \STATE mark(chunk, ingoing) 
    \ENDIF
    \ENDWHILE
  \end{algorithmic}
  \end{algorithm}

%
%

Figure 4 depicts the steps of Algorithm 1, in the lower part of FIGURE 2, source I.P. address and I.P. address range of SINET 
is translated to a 32-bit sequence.
Then, bit matching is done by a minus operation.
In the implementation, we use CUDA Thrust template library discussed in the following section.


\begin{figure}[ht]
\centering
\includegraphics[scale=1.5]{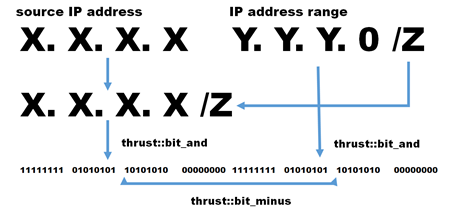}
\caption{Ingoing/outgoing session discrimination by bitmasking.}
\end{figure}

\subsection{TASK DECOMPOSITION}

If we want to transform code into a concurrent version, there are two ways. The first one is data decomposition, in which the program copes with a large collection of data and can compute every chunk of the data independently. The second one is task decomposition, in which the process is partitioned into a set of independent tasks that threads can execute in any order. Data decomposition has some drawbacks. For example, the size of divided session data files varies according to the situation in which the data is retrieved. The amount of session data changes ranging from 1,200,000,000 to 1,600,000,000. Besides, each divided process is NOT homogeneous. Four threads are working based on task-based decomposition and can handle massive threads for large session data files without considering tidy parameter selection in adopting data decomposition. Assume that we have 4 (the number of GPUs) threads, and each thread and each thread is associated with one session data file chunk. By doing this, Our system can take advantage of dynamic scheduling for coping with a variety of kinds and sizes of divided session data files. It consists of two thread sets: a master thread that enqueues the session data file's name and a worker thread that processes each session data file. The master thread enqueues the list of chunks of a session data file and passes it to the worker thread. More specifically, the master thread traverses the session data file directory and enqueues the file name. When the queue is full, the master thread waits until the worker thread processing packets consumes a file name and removes it from the queue.

\section{Two-stage Map-Reduce with tiling pattern}
We use a tiling pattern for two-stage Map Reduce, as shown in Figure 5.
Tiling is dividing a process into a set of parallel tasks of a suitable granularity.
Tiling can include multiple steps on a smaller part of a problem instead of running each step on the whole problem one after the other.
Tiling can lead to dramatic performance increases when a whole problem does not fit in the cache. 
Tiling can be applied using either a static decomposition or a divide-and-conquer strategy under programming models supporting fork-join.

\subsection{Map}

The basic concept of Map is (1) taking a collection of data and (2) associating a value with each item in the collection. Consequently, a collection of key-value pairs is generated by matching each input element with some related value. Also, the number of collections by mapping operation should be equal to the number of input data items in the collection. In the view of concurrency, it is important that pairing up keys and values are independent for each input in the collection.

\subsection{Reduce}

The basic concept of reducing is reducing or merging several different outputs from the map phase into a single output. Reduce operation yields the representative data required to produce the answer to the batch computation under processing by reducing the data from the data item. Similar to the map phase, a range of input should be equal to the one of output. Besides, the reduce phase can be repeated as many or as few times as required in order to yield the output down to a single value over the entire data set. In the view of concurrency, coordinated batch processing is a typical example because it can frequently happen regardless of the number of inputs split up. In this sense, it is similar to join, grouping together the parallel output of different batch operations.


\begin{figure}[ht]
\centering
\includegraphics[scale=0.7]{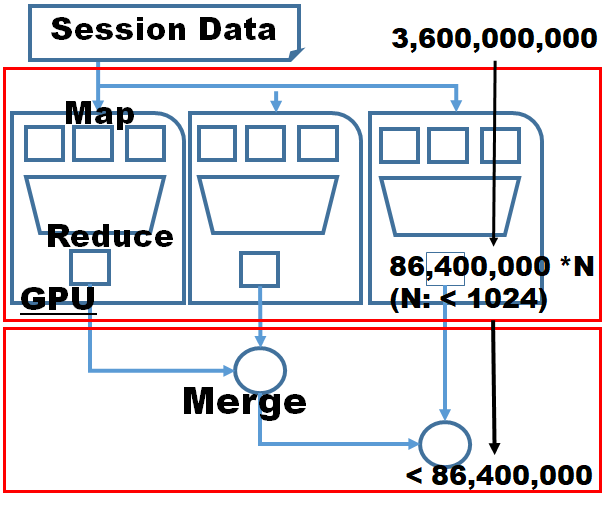}
\caption{Tiling pattern for two-stage map-reduce of GPU and CPU.}
\end{figure}

\subsection{Pairwise Reduction Pattern with GPU}
A common way to accomplish parallel addition using GPGPU is pairwise reduction. In pairwise reduction, a chunk contains a pair of elements (key-value). A thread sums two elements to yield one partial result. These intermediate results are stored in-place in the original input vector. Then, new values (sums) are used to input for summing in the next generation. For each iteration, the number of input values halves, which results in a final sum figured out when the length of the output vector reaches one. The parallel reduction is one of the most popular parallel patterns. The output of histogramming is Map such as <timestamp, count> and <timestamp, bytes>. Once the session data is loaded into device memory, histogramming can be executed on GPU. In this step, we use the CUDA Thrust template library. Thrust is a C++ template library for CUDA based on the Standard Template Library (STL). Thrust allows you to implement high-performance parallel applications with minimal programming effort through a high-level interface that is fully interoperable with CUDA C. Thrust provides a rich collection of data-parallel primitives such as scan, sort, and reduce, which can be composed together to implement complex algorithms with concise, readable source code. By describing your computation in terms of these high-level abstractions, you provide Thrust with the freedom to select the most efficient implementation automatically. As a result, Thrust can be utilized in rapid prototyping of CUDA applications, where programmer productivity matters most, and in production, where robustness and absolute performance are crucial. We use cudaSetDevice(i), which sets the device as the current device (i) for the calling host thread. The key point here is that GPU device I.D. is corresponding to thread I.D. by taking the argument of cudaSetDevice for thread\_id. The result\_A and d\_B are stored in local memory for each thread. As shown in the upper side of Figure 4, CUDA Trust API is invoked in each Pthread in parallel. At line 8, cudaSetDevice is invoked to assign GPU to the reduction operation. We use cudaSetDevice(i), which sets the device as the current device for the calling host thread. The key point here is that GPU device I.D. corresponds to thread I.D. by taking the argument of cudaSetDevice for thread\_id. The results d\_A and d\_B are stored in local memory for each thread.

\subsection{Merge Scatter Pattern with CPU}
The merge scatters pattern, associative and commutative operators are provided for merging elements in case of a collision. With the nature of this pattern, scatter could occur in any order. Therefore, both associative and commutative properties are required. An example that uses addition as the merge operator is shown in Figure 4. It is straightforward to adopt a merge scatter pattern to implement histograms with the adding operation. The interval length of aggregation of our system is a millisecond. Approximately, the granularity of histogramming is around 86,000,000 (60 * 60 * 24 * 1000 = 86,400,000). It is difficult to evade lock contention to store 86,000,000 key-value into a hash map in parallel from our experience. We give up using the concurrent hash map, which is affected by lock contention. Instead, key-value can be represented by using the defining namespace such as X1<timestamp>, X1<count> and X2<timestamp> and X2<bytes>. Containers provided by Intel TBB offer a much higher level of concurrency via one or both of the following methods:

Multiple threads operate on the container by locking only those portions they really need to lock. 
As long as different threads access different portions, they can proceed concurrently. Different threads account for and correct for the effects of other interfering threads. 

Notice that highly-concurrent containers come at a cost. They typically have higher overheads than regular STL containers. Operations on highly-concurrent containers may take longer than for STL containers. Therefore, using highly-concurrent containers when the speedup from the additional concurrency enables outweighs their slower sequential performance. A concurrent\_vector $ <T> $ is a dynamically growable array of T. It is safe to grow a concurrent\_vector while other threads are also operating on elements of it or even growing it themselves.

\section{Experimental Result}

\subsection{Dataset Description}

The PA-7000 Series is powered by a scalable architecture for the purpose of applying the appropriate type and volume of processing power to the key functional tasks of networking, security, and management.

\begin{table}[htb]
  \caption{PA-7080 data description}
  \begin{tabular}{{|l|c|r|}} \hline
    No & item name & value \\ \hline \hline
    1 & capture\_time & 2018/01/01 00:00:00.000   \\ \hline 
    2 & generated\_time &   2018/01/01 00:00:00 \\ \hline 
    3 & start\_time & 2018/01/01 00:00:00   \\ \hline 
    4 & elapsed\_time & 3  \\ \hline 
    5 & source\_ip & xxx.xxx.xxx.xxx \\ \hline 
    6 & source\_port & 0   \\ \hline 
    7 & src\_country\_code & JP  \\ \hline 
    8 & destination\_ip & yyy.yyy.yyy.yyy \\ \hline 
    9 & destination\_port & 0  \\ \hline 
    10 & dest\_country\_code & NA  \\ \hline 
    11 & protocol & NA  \\ \hline 
    12 & application & NA  \\ \hline 
    13 & subtype & NA  \\ \hline 
    14 & action &  NA \\ \hline 
    15 & session\_end\_reason & NA \\ \hline 
    16 & repeat\_count & 0  \\ \hline 
    17 & category & NA \\ \hline 
    18 & packets & 0  \\ \hline 
    19 & packets\_sent & 0  \\ \hline 
    20 & packets\_received & 0  \\ \hline 
    21 & bytes & 0  \\ \hline 
    22 & bytes\_sent & 0  \\ \hline 
    23 & bytes\_received & 0  \\ \hline 
    24 & device\_name & NA  \\ \hline     
  \end{tabular}
\end{table}

\begin{table}[htb]
  \caption{SINET session volume on Feb 2021}
  \begin{tabular}{{|l|c|r|}} \hline
    Date & sessions count & volume \\ \hline \hline
    02/19/2021(Fri) & 1632300495 &  633GB    \\ \hline 
    02/20/2021 (Sat) & 1328270546 &  514GB \\ \hline 
  \end{tabular}
\end{table}

The Session data format is shown in Table 1.
No.1 - 9 is concerned about TCP/IP packet header. NO 19-23 is retrieved to generate statistics.
Particularly, No.12 (application) and No.17 (category) are inspected in detail. 

Concerning traffic volume, we observe about 1.5-1.7 billion sessions per day on weekdays.
On weekends or holidays, we observe 1.2-1.4 billion sessions per day.
Table 2 shows the session data volume on 02/19/2021 and 02/20/2021 observed in SINET.

\subsection{Performance Measurement}

In the experiment, we use a rack server of Dell(TM) PowerEdge(TM) PE C4140 with 96 core CPUs and 4 GPUs. CPU is Intel(R) Xeon(R) Platinum 8160 CPU @ 2.10GHz. Our PE C4140 has 1.5TB memory. GPU is Tesla V100 PCIe 32GB (NVIDIA Corporation GV100GL),

Table 3 shows Elapsed time of session data ranging from 20,000,000 to 40,000,000 lines.
In the best case, a single GPU is 25.4 times faster than a single GPU at line 4 with data chunk size 16 G.B. and 4GPU.
Thrust API is fully asynchronous, and the contention may have occurred when we don't apply multi GPU.
Single GPU can hardly handle no more than 36 G.B. session data.

In summary, with data chunk size ranging from 16GB to 24GB, multi GPU (4 GPUs) are faster than single GPU 
by x7.8 (16GB, 3 GPUs), x25.4 (16GB, 4 GPUs), x13.07 (24GB, 3 GPUs) and x10.44 (24GB, 4 GPUs).
Concerning data chunk size of more than 36 G.B., a single GPU of Tesla V100 cannot handle the data chunk.

\begin{table}[htb]
  \caption{Elapsed time of session data ranging from 20,000,000 to 40,000,000 lines}
  \begin{tabular}{{|l|l|c|r|}} \hline
    1 GPU & 4 GPU & Speed Up \\ \hline \hline
    \multicolumn{2}{|c|}{chunk size: 4GB (20000000 lines) * 4 = 16GB} & \multicolumn{1}{c|}{} \\ \hline
    \multicolumn{2}{|c|}{Elaspsed time (sec) } & \multicolumn{1}{c|}{} \\ \hline
    single GPU &  4 GPUs  &    \\ \hline 
    1. 0.163212162 &  0.169472176  &    \\ \hline 
    2. 2.017158832 & 0.169368701 & \\ \hline
    3. 1.266480964 & 0.169198073 & x7.8 \\ \hline
    4. 4.328394731 & 0.172328982 & x25.4 \\ \hline  \hline
    \multicolumn{2}{|c|}{Data size: 6GB (30000000 lines) * 4 = 24GB} & \multicolumn{1}{c|}{}\\ \hline
    \multicolumn{2}{|c|}{Elaspsed time (sec) } & \multicolumn{1}{c|}{} \\ \hline
    single GPU &  4 GPUs  &    \\ \hline 
    5. 0.25455562 & 0.266662563 & \\ \hline
    6. 1.09780012 & 0.254225929 & \\ \hline
    7. 3.405360669 & 0.263795581 &  x13.07 \\ \hline
    8. 2.618192345 & 0.254559077 &  x10.44 \\ \hline  \hline
    \multicolumn{2}{|c|}{Data size: 9GB (40000000 lines) * 4 = 36GB} & \multicolumn{1}{c|}{}\\ \hline
    \multicolumn{2}{|c|}{Elaspsed time (sec) } & \multicolumn{1}{c|}{} \\ \hline
    single GPU &  4 GPUs  &    \\ \hline 
    10. 0.334865648 & 0.344755277 & \\ \hline
    11. out of memory & 0.337464371 & \\ \hline
    12. out of memory & 0.335592131 & \\ \hline
    13. out of memory & 0.33519815 & \\ \hline
  \end{tabular}
\end{table}

\subsection{All Sessions of SINET}

We first report the temporal pattern of the SINET ingoing/outgoing traffic patterns. 
Figures 6 and 7 show the traffic volume of session data.
Session data is grouped into one-hour frame bins. 
In ingoing traffic, we observe a clear diurnal pattern from 01/25/2021 to 01/29/2012.
This period is the weekday of the academic campus. 
On the other hand, 01/30/2021 and 01/31/2021 are weekend, which results in the decrease
of traffic and plateau.

\begin{figure}[ht]
\centering
\includegraphics[scale=0.3]{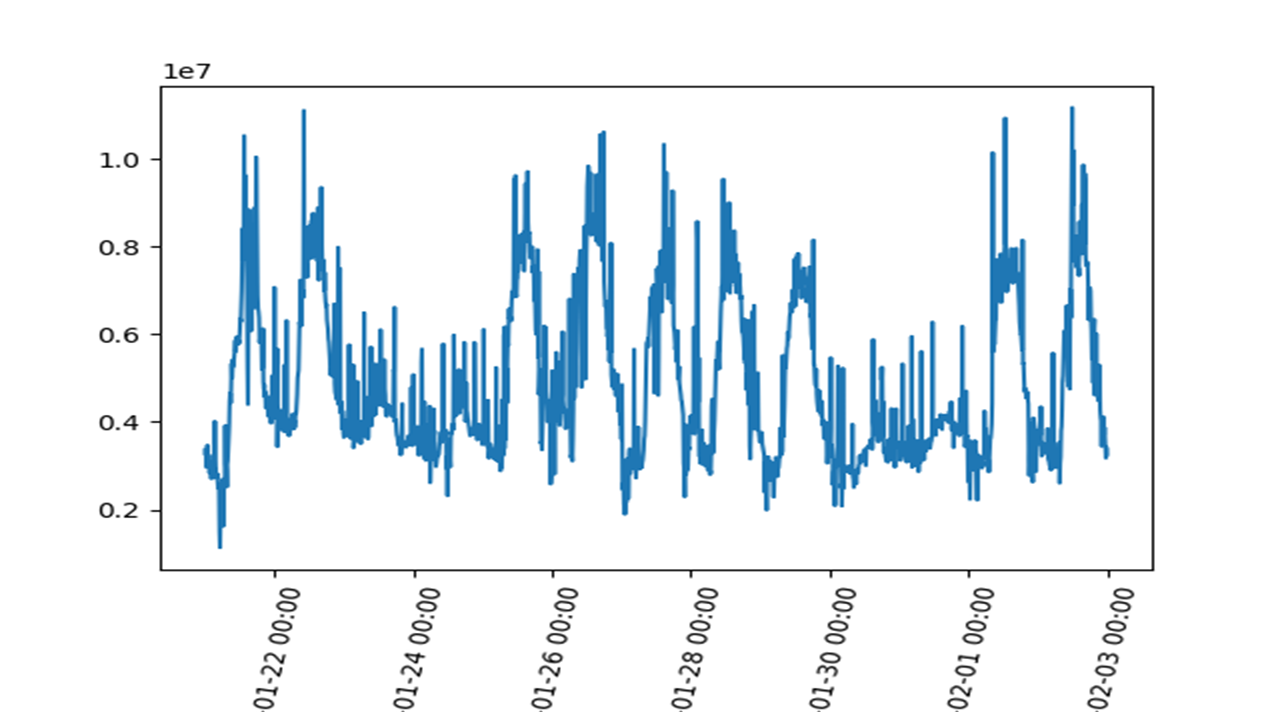}
\caption{SINET all ingoing sessions.}
\end{figure}



\begin{figure}[ht]
\centering
\includegraphics[scale=0.3]{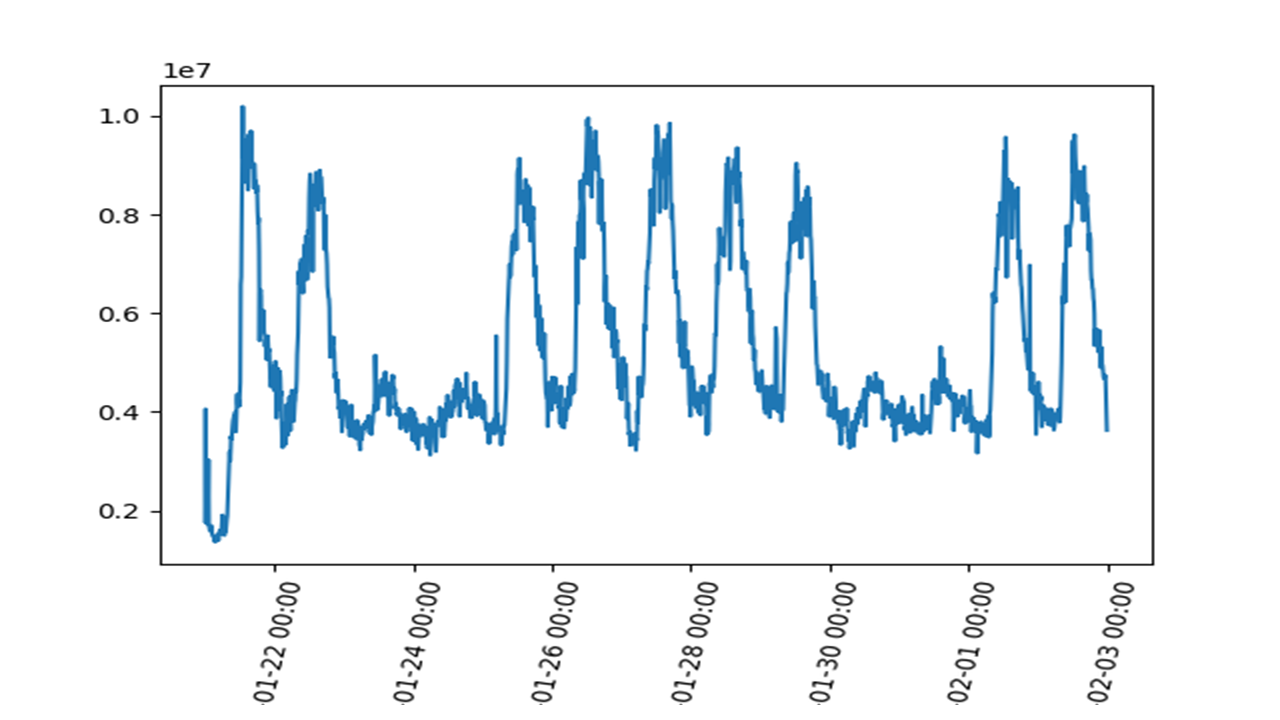}
\caption{SINET all outgoing sessions.}
\end{figure}

\subsection{AbuseIPDB}
Figure 8 and 9 shows the traffic volume of session data of the list of blacklist fetched from AbuseIPDB.
AbuseIPDB is a reputable website where system administrators can report information concerning I.P. address which abuses their resources.
AbuseIPDB has a backend database of shared I.P. addresses which involved in a malicious activity such as spam, exploit and DDOS attacks, etc.
In Figures 8 and 9, we randomly pick up 300 IP addresses by using AbuseIP blacklist API.
We observed a drastic traffic increase from 02/07/2021 to 02/08/2021.

\begin{figure}[ht]
\centering
\includegraphics[scale=0.3]{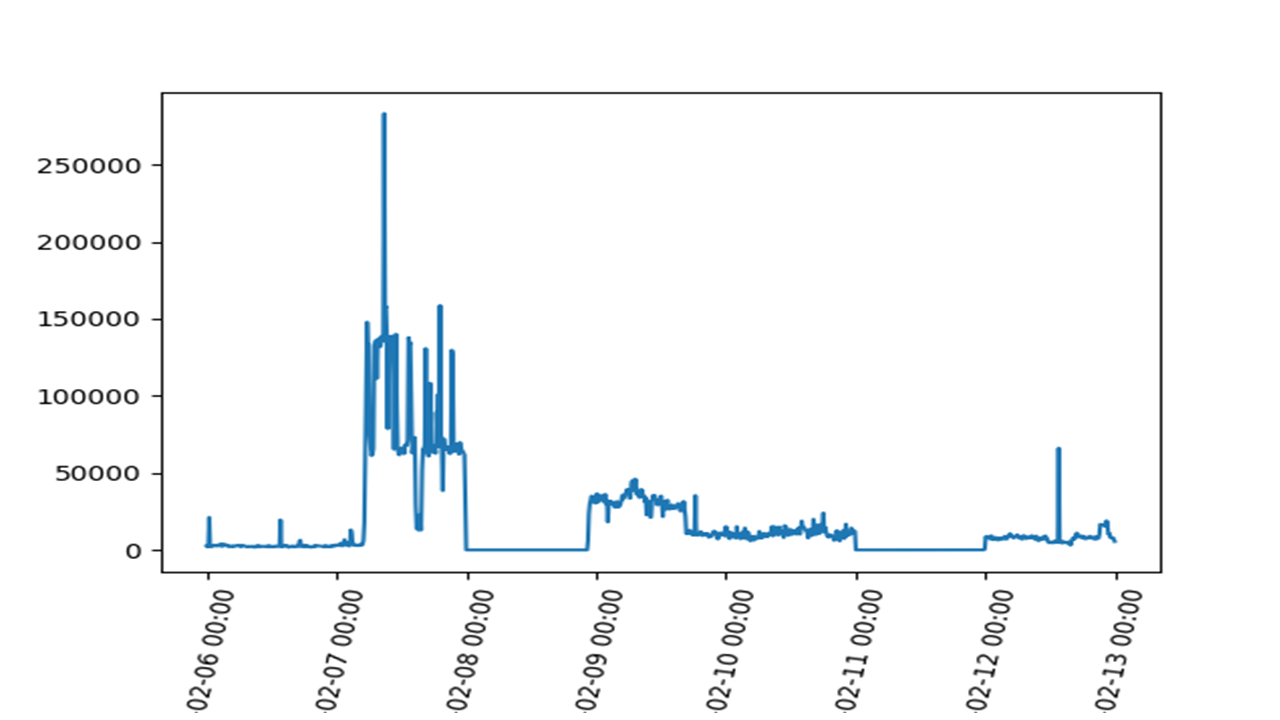}
\caption{All ingoing sessions randomly fetched by AbuseIPDB API.}
\end{figure}


\begin{figure}[ht]
\centering
\includegraphics[scale=0.3]{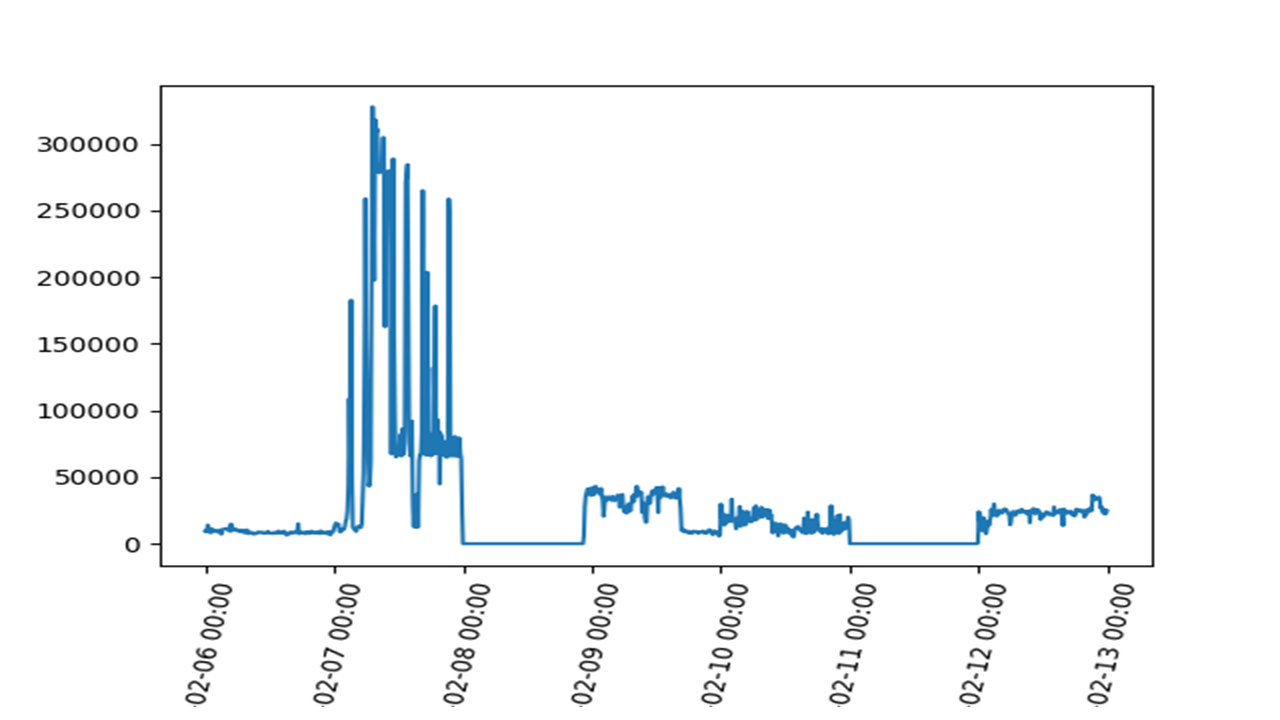}
\caption{All outgoing sessions randomly fetched by AbuseIPDB API.}
\end{figure}


\subsection{GRIZZLY STEPPE}
GRIZZLY STEPPE is Russian malicious cyber activity.  GRIZZLY STEPPE was designated by the Department Of Homeland Security (DHS) in October 2016. In December 2016, DHS and FBI released an analysis report detailing some of the tools and infrastructure used by adversaries of GRIZZLY STEPPE.
In Figures 10 and 11, we obtained 877 IP addresses from US-CERT report of GRIZZLY STEPPE.
We observed some traffic increases from 01/31/2021 to 02/13/2021. 
Among these spikes, the number of ingoing sessions on 02/05/2021 researched about 40,000 in 10 minutes.

\begin{figure}[ht]
\centering
\includegraphics[scale=0.3]{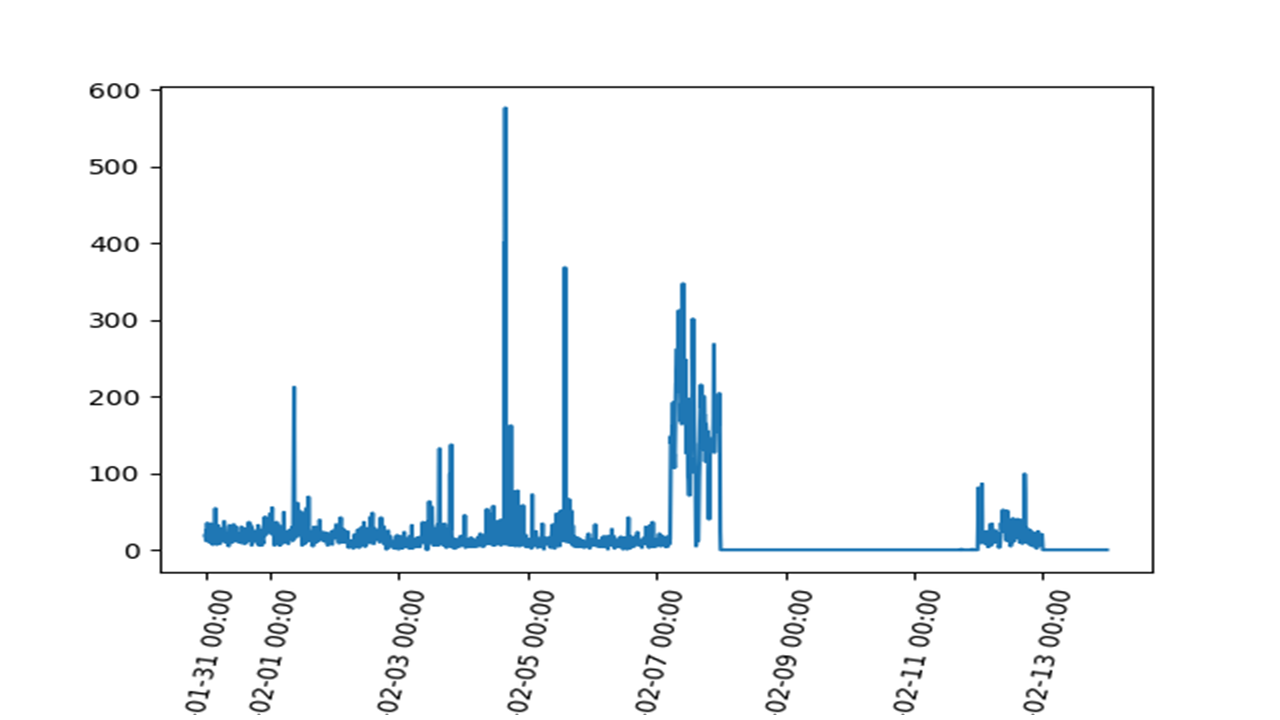}
\caption{Ingoing sessions of GRIZZLY STEPPE.}
\end{figure}


\begin{figure}[ht]
\centering
\includegraphics[scale=0.3]{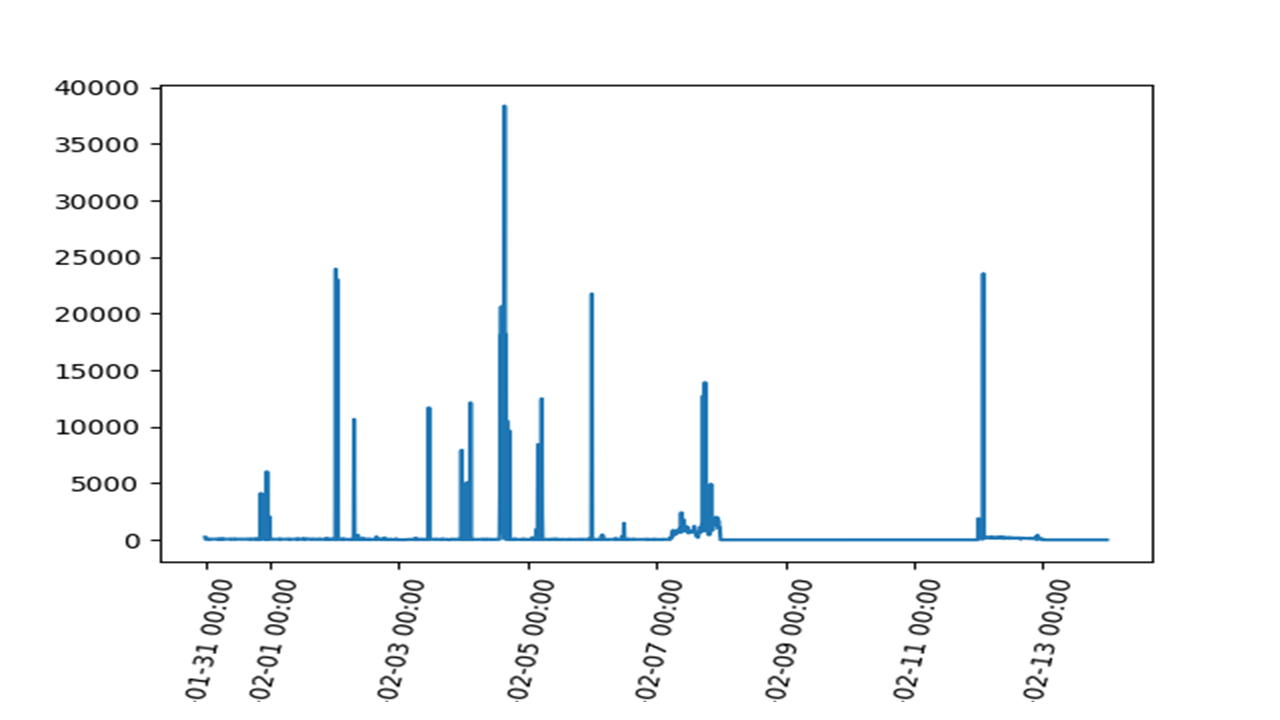}
\caption{Outgoing sessions of GRIZZLY STEPPE.}
\end{figure}


\section{Related work}
Fusco et al. propose a methodology for indexing a large amount of packet data per second by leveraging GPGPU \cite{GPU}.
Shredder \cite{shredder} leverages GPU for the design of a high-performance content-based chunking framework for supporting incremental storage and computation systems.
There have been many research efforts on understanding the traffic pattern of large-scale networks. 
Particularly, Li et al. propose an empirical analysis of mobile user access pattern of huge cloud storage service  \cite{Cloud}.
In \cite{Cellular}, Wang et al. show mobile traffic patterns of large-scale networks in an urban environment by gathering data in cellular towers.
Sandia National Laboratories (SNL) adopts Splunk for managing the Red Sky Supercomputer \cite{Splunk1}.
Bianca et al. adopt Splunk for optimizing data analysis with a semi-structured time-series database \cite{Splunk2}.
GPUs were initially designed for graphics rendering, but because of their cost-effectiveness, they were quickly adopted by the HPC
community for scientific computations \cite{Owens}. 
GPUs have also been used to accelerate functions such as pattern matching \cite{Shojania}, network coding \cite{Smith}. 

Asghari et al. \cite{Asghari} propose a systematic approach for providing comparative infection metrics from large-scale noisy sinkhole data of Conficker botnet. 
In \cite{Li}, a large dataset of 350 million HTTP request logs is analyzed for understanding user behavior of mobile cloud storage service. 
\cite{Xu} et al. report an empirical analysis for extracting and modeling the traffic patterns of 9600 cellular towers deployed in a metropolitan city. 
\cite{Bermudez} presents a characterization of Amazon's Web Services (AWS), which reveals that most of the content residing on EC2 and S3 is served by one Amazon data center located in Virginia.
\cite{Drago} et al. proposes a characterization of Dropbox by means of passive measurements of four vantage points in Europe, collected during 42 consecutive days.
In \cite{Jonker}, a new framework to enable a macroscopic characterization of attacks, attack targets, and DDOS protection is proposed.
\cite{Songbin} presents an analysis of online campus storage systems and data sharing services for more than 19,000 students and 500 student groups.
\cite{Miao} et al. present the large-scale characterization of inbound attacks towards the cloud and outbound attacks from the cloud using three months of NetFlow data in 2013 from a large cloud provider.

\section{Discussion}
In this paper, we give priority to handle huge session data over any other matter.
In 2009 researchers from Google wrote a paper entitled The Unreasonable Effectiveness of Data \cite{Alon}. 
According to this paper, if machine learning algorithm A using a messy dataset with a trillion lines can be highly 
effective in tasks, algorithm A is downright useless in coping with a clean dataset with a mere million lines.
If algorithm A does not work with a dataset comprising a million examples, our intuitive conclusion is that it does not work at all.
At the same time, it turns out that what is needed is a much bigger dataset and powerful computers and storage to process that much data.

\section{Conclusion}

In this paper, we present the multi-GPU accelerated pipeline for handle huge session data of SINET.
At each phase of the pipeline, we adopt a different kind of parallelisms such as thread-level parallelism and vector-level parallelism.
Our multi-GPU server is responsible for preprocessing session data between the workflow of ELK stack and Splunk.
On a multi-GPU server, we use a coordinated batch processing pattern for two kinds of procedure: discrimination and map-reduce.
In the step of discrimination, we use multi-GPU acceleration for CIDR matching with bitmasking. 
In the step of histogramming, we apply the design pattern of tiling for two stage map-reduce for rapid histogramming. 
In the experiment, we show a performance comparison of a single GPU and four GPUs. 
Also, we present some sample observations about SINET such as all SINET traffic, grizzely step, and AbuseIPDB.
With our multi-GPU-driven pipeline, we have succeeded in processing huge workloads of about 1.2 to 1.6 billion of session stream (500GB-650 G.B.) within 24 hours.
It can be concluded that multi-GPU acceleration makes it possible to handle huge session data of SINET.


\def\cprime{$'$} \def\cprime{$'$}


\bigskip

\end{document}